\documentstyle[pre,multicol,aps,epsfig,amssymb]{revtex}

\begin{document}

\title{Density profiles and surface tensions of polymers near
colloidal surfaces} 

\author {A.A. Louis$^1$, P.G. Bolhuis$^2$, E.J. Meijer$^2$ and
J.P. Hansen$^1$,} \address{$^1$Department of Chemisty, Lensfield Rd,
Cambridge CB2 1EW, UK} \address{$^2$Department of Chemical
Engineering, University of Amsterdam, Nieuwe Achtergracht 166, NL-1018
WV Amsterdam, Netherlands.}  \date{\today} \maketitle
\begin{abstract}
\noindent 
The surface tension of interacting polymers in a good solvent is
calculated theoretically and by computer simulations for a planar wall
geometry and for the insertion of a single colloidal hard-sphere.
This is achieved for the planar wall and for the larger spheres by an
adsorption method, and for smaller spheres by a direct insertion
technique.  Results for the dilute and semi-dilute regimes are
compared to results for ideal polymers, the Asakura-Oosawa
penetrable-sphere model, and to integral equations, scaling and
renormalization group theories.  The largest relative changes with
density are found in the dilute regime, so that theories based on
non-interacting polymers rapidly break down.  A recently developed
``soft colloid'' approach to polymer-colloid mixtures is shown to
correctly describe the one-body insertion free-energy and the related
surface tension.
\end{abstract}
\begin{multicols}{2}

\section{Introduction}

Binary mixtures of polymers and colloidal particles in various
solvents are the focus of sustained experimental and theoretical
efforts, both because they constitute a challenging problem in
Statistical Mechanics of ``soft matter'', and because of their
technological importance in many industrial processes.  One of the
most striking aspects of polymer-colloid mixtures, namely the
depletion interaction between colloids induced by non-adsorbing
polymer was recognized nearly 50 years ago\cite{Asak54}.  More
recently, the importance of the polymer depletant in determining the
phase behavior of the mixtures was realized\cite{LiIn75}, and much
recent experimental work was devoted to the phase
diagram\cite{Cald93,Ilet95,Poon99}, structure\cite{Weis99,Mous99},
interfaces\cite{deHo99}, and to the direct measurement of the
effective interactions\cite{Ohsh97,Verm98,Bech99}.  On the theoretical
side, most efforts have concentrated on impenetrable spherical
colloids, while various models and theoretical techniques have been
investigated for the description of the non-adsorbing polymer coils.
The models include non-interacting (ideal)
polymers\cite{Asak54,Meij94,Tuin00}, polymers represented as
penetrable-spheres\cite{Asak58,Lekk92,Loui99a,Brad00}, and interacting
polymers coarse-grained to the level of ``soft
colloids''\cite{Loui00,Bolh01,Bolh01a,Bolh01b}.  Monomer level
representations of polymer chains, like the self-avoiding walk (SAW)
model, appropriate in good solvent, have been considered within
polymer scaling approaches\cite{deGe79,Joan79,Tuin01}, renormalization
group (RG) theory\cite{Shaf99,Eise96,Hank99,Eise00,Maas01}, and fluid
integral equations\cite{Fuch00,Fuch01}.

While many effects for the simplest case of colloids mixed with
non-interacting polymer are quantitatively understood, the behavior of
the experimentally more relevant case of polymers with excluded volume
interactions is at best understood on a qualitative basis; a
quantitatively reliable theory is still lacking.  Clearly, to
construct such a theory for finite concentrations of colloidal
particles, one must first understand how interacting polymer coils
distribute themselves around a single spherical colloid of radius
$R_c$.  This problem is addressed in the present paper using a
combination of Monte Carlo (MC) simulations and scaling theories to
determine the key quantities, which are the monomer or center-of-mass
(CM) density profiles $\rho(r)$ of SAW polymers around a single
impenetrable sphere, as well as the resulting surface tension.  If
$R_g$ denotes the radius of gyration of the polymers, these quantities
clearly depend on the ratio $q=R_g/R_c$, which controls the curvature
effects.  The limit $q \rightarrow 0$, corresponding to a polymer
solution near a planar wall, will be examined first, before
considering the case of finite $q$.  The complete theory for the
opposite limit, $q \gg 1$, will be the subject of a future
publication, although we show some preliminary results here.
Throughout this work we focus on the dilute and semi-dilute
regimes\cite{deGe79,Doi86} of the polymers, where the monomer density
$c$ is low enough, for detailed monomer-monomer correlations to be
unimportant; the melt regime, where $c$ becomes appreciable, will not
be treated here.

The surface tension of a polymer solution surrounding a sphere is
macroscopically defined by considering the immersion of a single hard
colloidal particle into a bath of non-adsorbing polymer.  Because this
immersion reduces the number of configurations available to the
polymers, resulting in an entropically induced depletion layer around
the colloid, there is a free energy cost $F_1$ for adding a single 
sphere to the polymer solution which naturally splits into volume and
surface terms:
\begin{equation}\label{eq1.1}
F_{1} = \Pi(\rho) \frac{4}{3} \pi R_c^3 + 4 \pi R_c^2
\gamma_{s}(\rho).
\end{equation}
The first term in Eq.~(\ref{eq1.1}), describes the reversible work
needed to create a cavity of radius $R_c$ in the polymer solution.
Since the osmotic pressure $\Pi(\rho)$ of a polymer solution in the
dilute and semi-dilute regimes is quantitatively known as a function
of polymer concentration $\rho$ from RG calculations\cite{Shaf99},
this volume term is well understood.  The problem of a quantitative
description of a single colloid in a polymer solution thus reduces to
understanding the second term, which defines the surface tension
$\gamma_s(\rho)$, i.e.\ the free-energy per unit area that is directly
related to the creation of the depletion layer. It is customary
to relate the surface tension $\gamma_s(\rho)$ around a sphere to 
 the surface tension $\gamma_w(\rho)$ near  a planar wall, by
expanding in powers of the ratio $q=R_g/R_c$:
\begin{equation}\label{eq1.2}
\gamma_s(\rho) = \gamma_w(\rho) + \kappa_1(\rho) q + \kappa_2(\rho) q^2
+  {\cal
O}(q^3),
\end{equation}
which is expected to be most useful when $q$ is not too large.  The
coefficients $\kappa_{i}(\rho)$ control the curvature corrections.
They are analogous to the Tolman corrections in the macroscopic
case\cite{Tolm49,Rowl89}.

The paper is organized as follows: The case of a single plate or hard
wall immersed in a polymer solution is discussed in section II, where
we report results for density profiles $\rho(z)$ at various polymer
concentrations.  These density profiles define the reduced adsorption
$\hat{\Gamma}(\rho)$, from which the surface tension $\gamma_w(\rho)$
may ultimately be extracted.  These considerations are extended to
spherical colloids in section III, where simulation results for the
density profiles are reported for size ratios $q=0.67, 1.05$ and
$1.68$.  These data are then used to compute $\gamma_s(\rho)$ and the
$\kappa_i(\rho)$; limiting forms are extracted for the $\rho
\rightarrow 0$ and the semi-dilute regimes.  The results are compared
to the theoretical predictions for ideal polymers, for the penetrable
sphere model, and wherever applicable, to RG and integral equation
predictions. The limit of large $q$, where the expansion (\ref{eq1.2})
becomes less useful is also discussed.  For this limit we also report
on some preliminary direct simulation results for $F_1$ based on the
Widom insertion technique\cite{frenkelbook}.  Finally we show that the
``soft colloid'' paradigm has the correct thermodynamics of the single
colloid problem automatically built in.

\section{Density profiles and surface tension near a single wall}

A single hard wall in a bath of non-adsorbing polymers creates an
entropically induced depletion layer because the polymers have fewer
possible configurations near the wall.  To calculate these density
profiles we performed Monte-Carlo simulations of the popular self
avoiding walk (SAW) model on a cubic lattice.  Even though this model
ignores all chemical details of a real polymer system except the
excluded volume and polymer connectivity, it reproduces the scaling
behavior and many other physical properties of athermal polymer
solutions~\cite{deGe79,Doi86}.  For $N$ polymers of length $L$ on a
lattice of $M$ sites, the polymer density is given by $\rho = N/M$,
while the monomer density  is $c = L N /M$.  The polymers are
characterized by the radius of gyration, which scales as $R_g \sim
L^{\nu}$, where $\nu \approx 0.588$ is the Flory
exponent\cite{deGe79,Shaf99,Doi86}.  For densities $\rho$ less than the
overlap density $\rho^* = 1/(\frac43 \pi R_g^3)$ the system is in the
dilute regime, while for $\rho \geq \rho^*$, and $c \ll 1$, the system
is in the semi-dilute regime.  We use $L=500$ SAW polymers in our
simulations, which are expected to exhibit properties close to those
corresponding to the scaling limit $L \rightarrow \infty$.  Further
details of the simulation method and the model can be found
in\cite{Loui00,Bolh01,Bolh01a,Bolh01b}.  Note that a small correction
to these results must be applied\cite{Rg}!  Since our models are all
athermal, we set $\beta = 1/(k_B T) =1$.

\begin{figure}
\begin{center}
\epsfig{figure=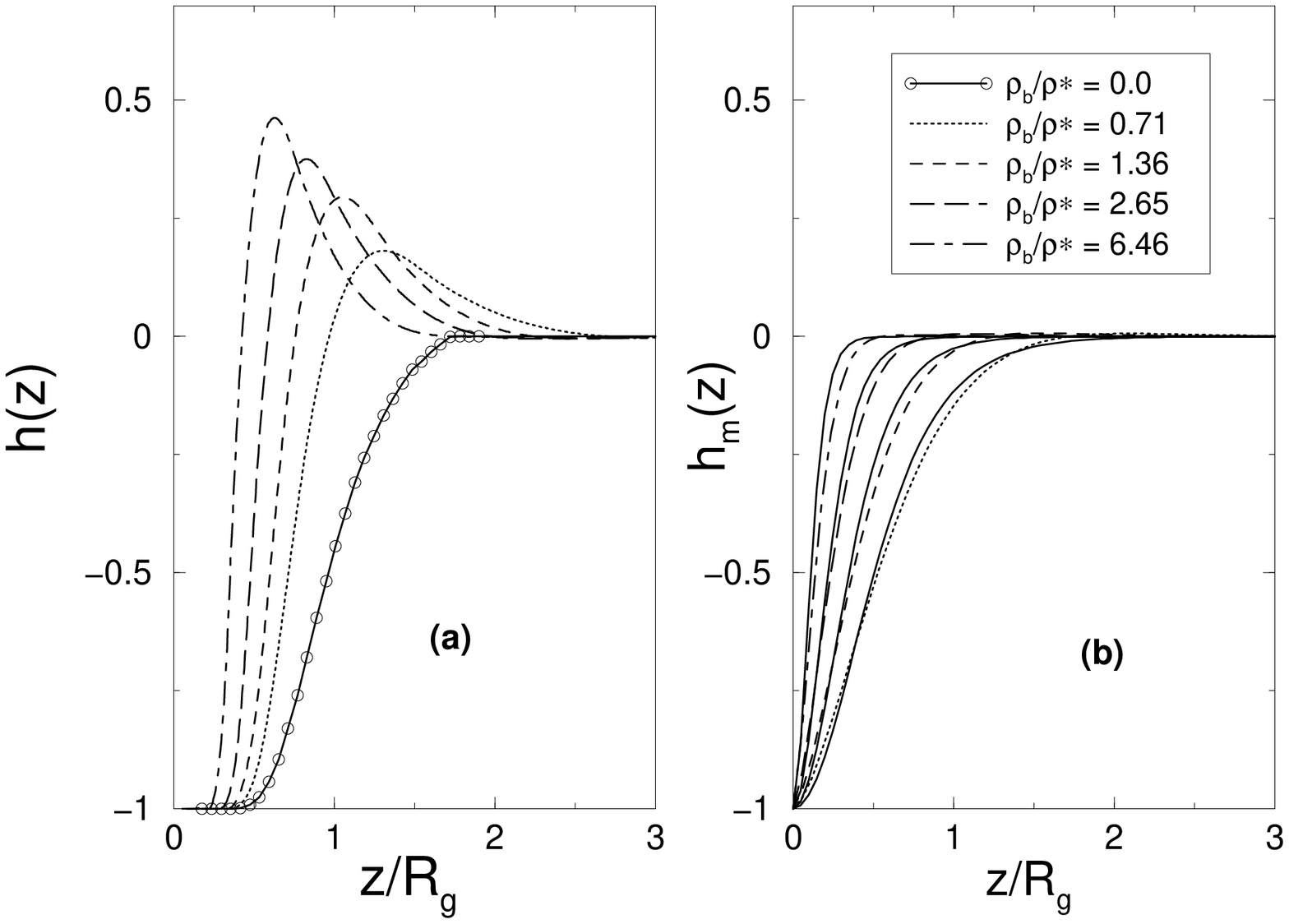,width=8cm}
\caption{\label{fig:hz} {\bf (a)} The wall-polymer CM profile
$h(z) = \rho(z)/\rho -1$ for $L=500$ SAW polymers at different
bulk concentrations.  {\bf (b)} The wall-monomer profile
$h_m(z)$ for the same bulk concentrations.  Both representations
result, by definition, in the same relative adsorptions.  The full
lines are a fit to the simple form $h_m(z) = \tanh^2(z/\hat{\Gamma}(\rho))
-1$}
\end{center}
\end{figure}

Examples of the depletion layer density profiles near a hard wall are
depicted in Fig.~\ref{fig:hz} for a polymer center of mass (CM) 
representation, as well as for a monomer representation.  Both
profiles have, by definition, the same reduced adsorption, defined
as\cite{Rowl89}:
\begin{equation}\label{eq2.3}
\hat{\Gamma}(\rho) = - \frac{1}{\rho}\frac{\partial
 ( \Omega^{ex}/A)}{\partial \mu} = \int_{0}^\infty h(z) dz,
\end{equation}
where $\Omega^{ex}/A$ is the surface excess grand potential per unit
area $A$.  $h(z) = \rho(z)/\rho -1$, with $\rho(z)$ the CM density
profile of the polymer coils a distance $z$ from the surface. In the
monomer representation one should replace $\rho$ by $c =  LN/M$ and
$h(z)$ by the monomer profile; the two reduced adsorptions are equal
and measure the reduction in the number of chains near the surface.

  In the low-density limit an RG calculation based on a first order
$\epsilon$-expansion gives $\hat{\Gamma}(0) \approx - 1.074
R_g$\cite{Hank99,gamma0}, which is slightly less than 
\begin{equation}\label{eq2.3b}\hat{\Gamma}^{id} =
 2R_g/\sqrt{\pi} \approx - 1.128 R_g,
\end{equation} the density independent result for an ideal polymer
with the same size $R_g$\cite{Eise96} (but larger $L$ due to the
different scaling of the radius of gyration).  This reflects the fact
that for a given $R_g$, the polymer-polymer interactions reduce the
size of the depletion layer, an effect which becomes more pronounced
with increasing density, see e.g.\ Fig.~\ref{fig:hz}.

 For the semi-dilute regime, de Gennes has proposed an approximate
expression for the monomer profile near a wall, $h_m(z) =
\tanh^2(z/\xi(\rho)) -1$, where $\xi(\rho)$ is the correlation length
or blob size\cite{deGe79}. If we identify $\xi(\rho)$ with
$-\hat{\Gamma(\rho)}$ then, as shown in Fig~\ref{fig:hz}, this form
provides a fairly accurate fit to our simulation results.  Since
$\xi(\rho) \sim \rho^{-\nu/(3\nu -1)} \sim \rho^{-0.770}$ in the
semi-dilute regime, this implies that the density profiles should
become more narrow with increasing density, a trend clearly seen in
Fig.~\ref{fig:hz}.

\begin{figure}
\begin{center}
\epsfig{figure=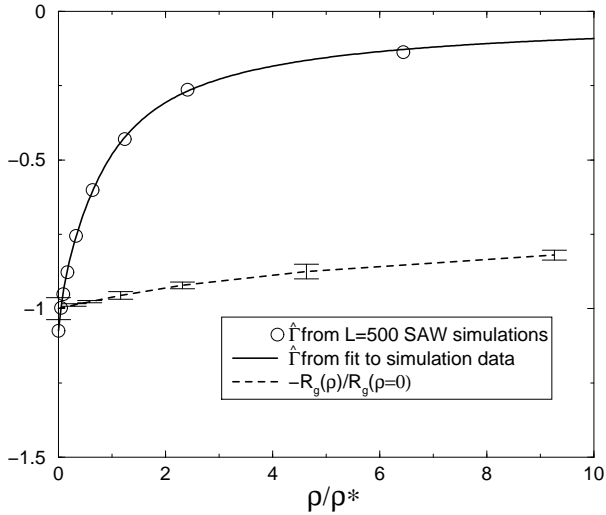,width=8cm}
\caption{\label{fig:Gamma-wall-fit} Relative adsorption
$\hat{\Gamma}(\rho)$, in units of $R_g$, versus density.  Circles
denote direct computer simulations of $L=500$ SAW polymers near a
single hard wall, and the line denotes the simple fit with the correct
scaling behavior, given by Eq.~(\protect\ref{eq2.5}).  Also shown is
the density dependence of the radius of gyration.  In the semi-dilute
regime $\hat{\Gamma} \approx -\xi \sim \rho^{0.770}$, while
$R_g(\rho)/R_g(\rho=0) \sim \rho^{0.115}$}
\end{center}
\end{figure}

 From the density profiles of Fig.~\ref{fig:hz}, we can derive the
adsorption at different densities using Eq.~(\ref{eq2.3}). These are
shown in Fig.~\ref{fig:Gamma-wall-fit}, together with a simple fit
constrained to give the expected $\rho=0$ value,  and the correct
scaling behavior in the semi-dilute regime where $-\hat{\Gamma}(\rho)
\approx \xi(\rho) \sim \rho^{-0.770}$, namely
\begin{equation}\label{eq2.5}
\hat{\Gamma}(\rho) =- 1.074 R_g\left( 1 + 7.63\frac{\rho}{\rho^*} +
14.56(\frac{\rho}{\rho^*})^3\right)^{-(0.2565)}.
\end{equation}
Throughout this paper the value of the radius of gyration is
conventionally chosen as that appropriate for an isolated polymer,
i.e. $R_g = R_g(\rho=0)$.  However, as the polymer concentration
increases, the measured $R_g(\rho)$ will decrease with density as
shown in Fig.~\ref{fig:Gamma-wall-fit}.  In the semi-dilute regime
this scales as $R_g(\rho)/R_g(\rho=0) \sim \rho^{(2 \nu -1)/(6 \nu -
2)}\approx \rho^{-0.115}$\cite{deGe79}, which decreases much more
slowly with density than the correlation length $\xi(\rho)$ or the
relative adsorption $\hat{\Gamma}(\rho)$.  In fact at $\rho/\rho^*=1$,
the crossover from the dilute to the semi-dilute regimes,
$\hat{\Gamma}(\rho)$ has dropped to $59\%$ of its $\rho \rightarrow 0$
value, while $R_g(\rho)$ has only changed by a few $\%$.  The largest
rate of relative change in the adsorption is therefore found in the
dilute regime, suggesting that theories based on the $\rho \rightarrow
0$ limit may start to break down well before the semi-dilute regime is
reached.  The border between the dilute and semi-dilute regimes is not
sharp.  For the semi-dilute regime, the asymptotic forms derived by
scaling theories appear to be reached at a lower density for
$\hat{\Gamma}(\rho)$ than for $R_g(\rho)$\cite{Bolh01}.

One route to calculate the surface tensions from the density profiles
is to use an extension of the Gibbs adsorption equation to express the
surface tension near a single wall in terms of the relative adsorption
and the equation of state:
\begin{equation}\label{eq2.2}
\gamma_w(\rho) = \frac{ \partial \Omega^{ex}}{\partial A} = -
\int_{0}^{\rho} \left( \frac{\partial \Pi(\rho')}{\partial \rho'}\right)
\hat{\Gamma}(\rho') d\rho'.
\end{equation}
The derivation of this equation can be found, for example,
in\cite{Mao97,Tuin01}.  By performing one  integration by parts
w.r.t.\ density, Eq.~(\ref{eq2.2}) can also be expressed as:
\begin{equation}\label{eq2.4}
\gamma_w(\rho) = -\Pi(\rho) \hat{\Gamma}(\rho) +
\int_{0}^{\rho} \Pi(\rho')\left( \frac{\partial
\hat{\Gamma}(\rho')}{\partial \rho'}\right) d\rho',
\end{equation}
  The first term in this equation takes the form of a pressure times a
length.  For ideal polymers, where $\hat{\Gamma}(\rho)$ is
independent of density\cite{Asak54,Bolh01}, this term completely
describes the surface tension of the depletion layer.  It is just the
(entropic) free energy cost per unit area of creating a cavity of
volume $\hat{\Gamma}(\rho) A$.  The second (positive) term is
therefore only relevant if there are polymer-polymer interactions.

\begin{figure}
\begin{center}
\epsfig{figure=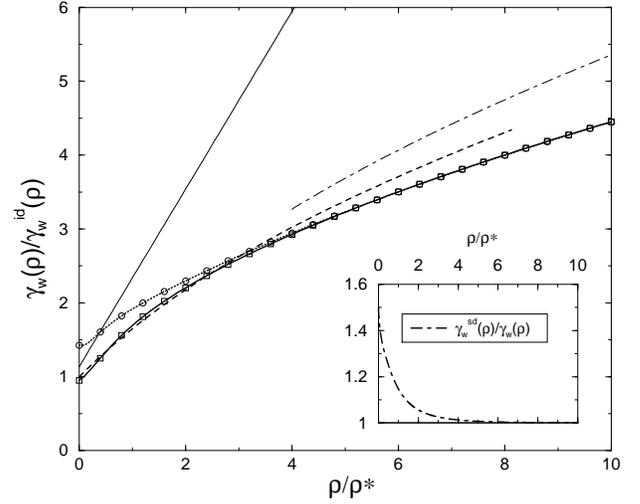,width=8cm}
\caption{\label{fig:gamma-wall} Polymer-wall surface tension
$\gamma_w(\rho)$ divided by $\gamma_w^{id}(\rho)$. The full lines with
square symbols are for interacting polymers and were calculated with
Eq.~(\protect\ref{eq2.4}), while the dotted line with circle symbols
denotes the simpler expression of Eq.~(\protect\ref{eq2.7}) which is
only valid in the semi-dilute scaling regime, where $\gamma_w \sim
\rho^{1.539}$.  Also shown are two recent RG
results\protect\cite{Maas01}: the dashed line denotes the renormalized
tree expansion, while the dot-dashed line denotes the asymptotic limit
for an $\epsilon$-expansion.  The solid line is from
Eq.~(\ref{eq2.8}), a result derived from a recent PRISM
calculation\protect\cite{Fuch01}.  The inset shows the ratio of the
full and simplified expressions for $\gamma_w(\rho)$. They coincide
for higher densities but in the low density limit, the semi-dilute
scaling expression overestimates the true surface tension by a factor
$1.5$.}
\end{center}
\end{figure}

We have previously calculated the equation of state for $L=500$ and
$L=2000$ SAW polymers\cite{Bolh01,Bolh01b}, both of which are well
described by analytic RG expressions\cite{Ohta82}.  Using this for
$\Pi(\rho)$ together with the fit to the relative adsorption from
Eq.~(\ref{eq2.5}), we can now use Eq.~(\ref{eq2.4}) to calculate the
surface tension of a solution of polymers in good solvent near a
single wall.  Our results are shown in Fig~\ref{fig:gamma-wall}.  In
the low density limit the surface tension reduces to the same
functional form as for ideal polymers, i.e.\ $\gamma_w(\rho) = -
\Pi(\rho) \hat{\Gamma}(\rho)$, so that $\lim_{\rho/\rho^* \rightarrow
0}  \gamma_w(\rho) \approx 1.074 R_g \rho$.  Note that for all
but the lowest densities, the surface tension is considerably larger than
the ideal polymer result $\gamma_w^{id}(\rho) = 2 \rho R_g
/\sqrt{\pi}$.  The surface tension for interacting polymers increases
more rapidly with increasing density both because $\Pi(\rho)$
increases faster than $\hat{\Gamma}(\rho)$ decreases in the first
term of Eq.~(\ref{eq2.4}), and because the second term, which is
absent for non-interacting polymers, increases with density as well.


Further simplifications occur in the semi-dilute regime. For example,
when the scaling forms for the osmotic pressure, $\Pi \sim \rho^{3
\nu/(3 \nu -1)}$, and for the reduced adsorption, $\hat{\Gamma} \sim
\rho^{-\nu/(3 \nu -1)}$, are used in Eq.~(\ref{eq2.4}), then the
integral in the second term can be easily performed and turns out to
be exactly half the first term, a result that is independent of the
value of the exponent $\nu$.  The surface tension therefore takes on a
very simple form:
\begin{equation}\label{eq2.7}
\gamma_w^{sd} (\rho) = -\frac{3}{2} \Pi(\rho) \hat{\Gamma}(\rho)
\sim \rho^{2 \nu/(3 \nu -1)} \approx \rho^{1.539}.
\end{equation}
As shown in Fig~\ref{fig:gamma-wall}, this expression works remarkably
well for larger densities into the semi-dilute regime.  Deviations do
occur for the dilute regime where Eq.~(\ref{eq2.7}) overestimates the
surface tension by a factor $1.5$ for $\rho \rightarrow 0$, as
demonstrated in the inset of Fig.~\ref{fig:gamma-wall}.

  In a recent publication Maassen, Eisenriegler, and
Bringer\cite{Maas01} have used the renormalized tree approximation to
derive a surface tension which compares  well with our
results, as shown in Fig.~\ref{fig:gamma-wall}.  A similar asymptotic
RG $\epsilon$-expansion compares slightly less well. The difference
between the two approximations gives an estimate of the error in the
RG approach.  It should be kept in mind that our simulation approach
also incurs small errors through the use of the fitted form of
$\hat{\Gamma}(\rho)$, and because we use polymers of a finite
length.

 Fuchs and Schweitzer\cite{Fuch01} recently applied the polymer
reference interaction site model (PRISM) approach to polymer-colloid
mixtures.  In the limit of low colloid density, a number of analytic
results can be derived for the insertion free energy $F_1$, from
which the surface tension can be extracted by using Eq.~(\ref{eq1.1})
and Eq.~(\ref{eq1.2}):
\begin{equation}\label{eq2.8}
\gamma_w^{PRISM}(\rho) = 1.279 \rho R_g (1 + 1.06 \frac{\rho}{\rho*}).
\end{equation}
Here we have used the PRISM results arising from local packing
information (see \cite{Fuch01} for details).  As can be seen in
Fig.~\ref{fig:gamma-wall} PRISM shows the correct qualitative but not
the correct quantitative behavior.  This is not surprising since
these results are based on a simplified PRISM model which has the
advantage of being analytically solvable, but the disadvantage of
exhibiting the wrong scaling behavior.  It would be interesting to
see how well PRISM compares for the numerically more complex case
where correct scaling behavior is included from the outset.

\section{Density profiles and surface tension around a hard sphere}

Having described the surface tension for a polymer solution near a
single hard wall, we next investigate the related problem of a polymer
solution near a single hard sphere (HS) of radius $R_c$.  As discussed in
the Introduction, adding a single HS to a polymer solution
reduces the number of configurations available to the polymers, and
results in a finite insertion free energy or chemical potential described by
Eq.~(\ref{eq1.1}).  Besides the configurations directly excluded by the
sphere of volume $\frac43 \pi R_c^3$, there are also configurations
excluded near the surface of the sphere, an effect which manifests
itself in an entropically driven depletion layer, just as was found
for the case of a hard wall.  However, the curvature of the sphere
leads to corrections to the planar surface tension, as described by
Eq.~(\ref{eq1.2}), i.e.\ the  surface tension $\gamma_{s}(\rho)$
 depends not only on the polymer density $\rho$, but also on
$R_c$ through the ratio $q=R_g/R_c$.

\subsection{Ideal polymers}
The free energy cost of inserting a single HS into a bath of
ideal polymers is know\cite{Eise96,Warren}:
\begin{equation}\label{eq13.2}
F_1^{id} = \frac{\rho}{\rho^*} \frac{1}{q^3} \left(1 + \frac{6
 q}{\sqrt{\pi}} +  3 q^2 \right) .
\end{equation}
By combining this result with Eq.~(\ref{eq1.1}), it follows that the
ideal polymer surface tension takes the form:
\begin{equation}\label{eq13.3}
  \gamma^{id}_s(\rho) =  \gamma_w^{id}(\rho)  +
\rho  R_g q
\end{equation}
The curvature corrections defined in Eq.~(\ref{eq1.2}) take on a
particularly simple form here, since $\kappa_1^{id}(\rho) = \rho
R_g$, and $\kappa_i=0$ for $i \geq 2$.  Note that this expression is
not simply an expansion in $q=R_g/R_c$; it is valid for all size
ratios.

In 1958 Asakura and Ooswawa\cite{Asak58} introduced a model where the
ideal polymers are approximated as inter-penetrable spheres of
radius $R_{AO}$.  This corresponds to approximating the true depletion
layer by a step-function. The free-energy of insertion of a single hard
sphere into a bath of AO particles can be easily calculated to be
\begin{equation}\label{eq13.4}
 F_1^{AO} = \eta_{AO} \frac{1}{q_{AO}^3} \left(1 + q_{AO} \right)^3.
\end{equation}
where $\eta_{AO} = \frac43 \pi \rho R_{AO}^3$ is analogous to
$\rho/\rho^*$, and we have defined the size ratio $q_{AO} =
R_{AO}/R_c$.  The surface tension is therefore given by
\begin{equation}\label{eq13.5}
 \gamma^{AO}_s(\rho) =  \gamma^{AO}_w(\rho) + \rho
 R_{AO} q_{AO} + \frac{\rho
 R_{AO}}{3}\left(q_{AO}\right)^2
\end{equation}
where $ \gamma_w^{AO}(\rho) = \rho R_{AO}$.  In this case the
curvature corrections have a very simple geometrical origin: The volume
of a spherical shell of width $R_{AO}$ with an inner radius of $R_c$
has a larger volume than that of a flat layer of width $R_{AO}$ and
area $4 \pi R_c^2$.  In part this is a matter of definition. For hard
particles one can also find instances in the literature where $R_c
+R_{AO}$ is taken as the radius of the Gibbs dividing surface.  The AO
model surface tension vanishes if it is defined in this way.

\begin{figure}
\begin{center}
\epsfig{figure=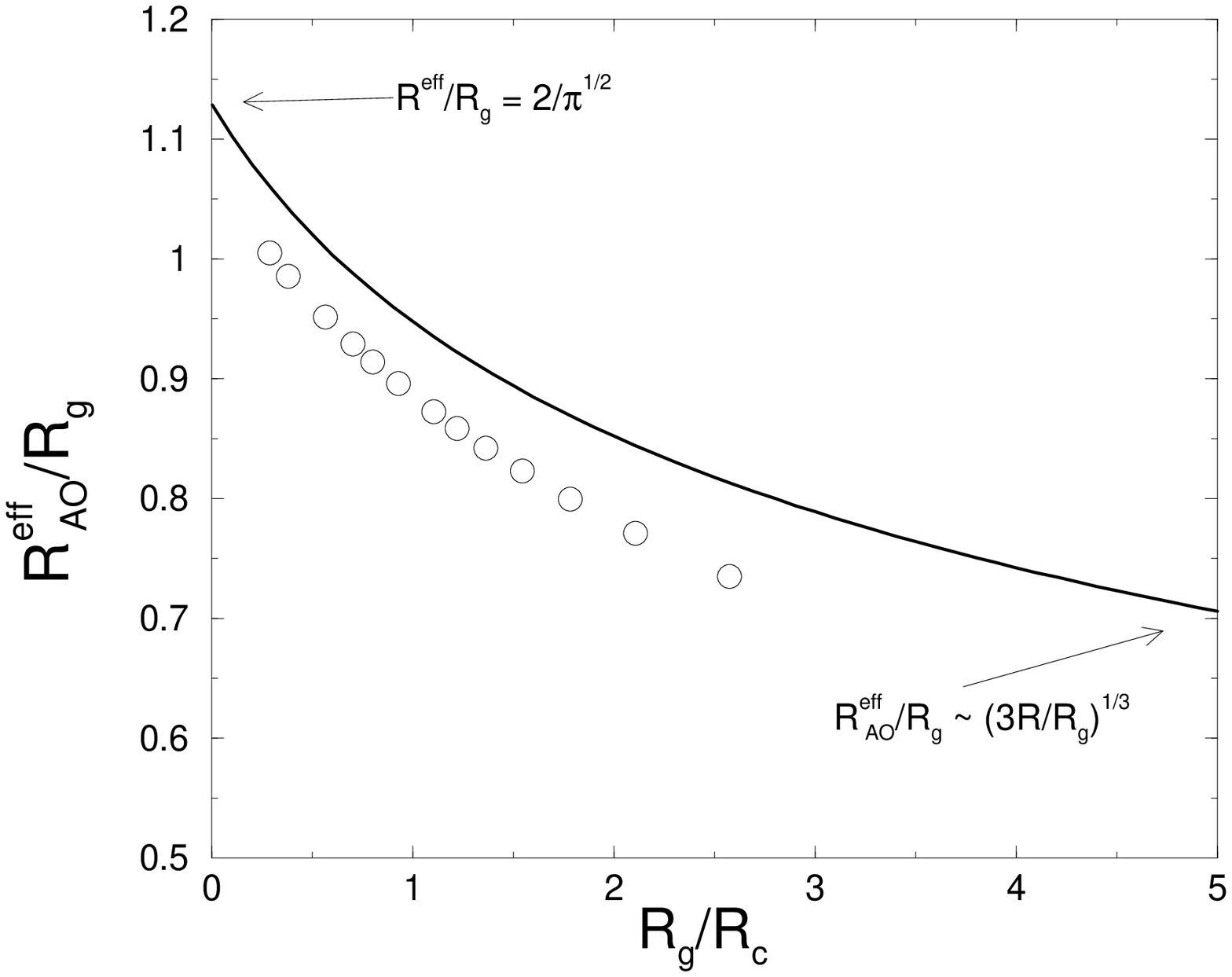,width=8cm}
\caption{\label{fig:EJM-Fig2} The effective AO radius $R_{AO}^{eff}$,
given by Eq.~(\protect\ref{eqA.4}), which would result in the same
surface tension for an AO fluid around a sphere of radius $R_c$ as
found for ideal polymers of size $R_g$. For infinite sphere size
(i.e.\ a wall) $R_{AO} = (2/\sqrt{\pi}) R_g$. As the relative sphere
size $R_c/R_g$ decreases this effective parameter decreases due to the
deformation of the polymers around a sphere.  The symbols denote
direct simulations $L=200$ ideal polymers on a lattice, taken from
reference\protect\cite{Meij94}.  The small differences are due to the
discrete nature of the lattice used in the simulations.  }
\end{center}
\end{figure}

If the prescription $R_{AO}= 2 R_g/\sqrt{\pi}$ is used to set the free
parameter in the AO model, then the surface tensions for the planar
wall is the same as that of ideal polymers.  However, this
prescription no longer holds for spheres immersed in a polymer
solution since the curvature corrections to the surface tension for
ideal polymers are not the same as those of the AO model.  Physically
this difference arises because the AO model assumes a fixed depletion
layer width $R_{AO}$ while the (ideal) polymers can deform around a
sphere, which leads to an effectively smaller depletion layer.  This
effect becomes progressively more pronounced with decreasing sphere size
$R_c/R_{g}$\cite{Meij94}.  An effective AO parameter which takes this
deformation effect into account can be derived by equating the two
surface tensions, Eq.~(\ref{eq13.3}) and Eq.~(\ref{eq13.5})
\begin{equation}\label{eqA.4}
\frac{R_{AO}^{eff}}{R_g} = \frac{1}{q} \left( \left(1 +
 \frac{6}{\sqrt{\pi}} q + 3 q^2 \right)^{1/3}
 -1 \right).
\end{equation}
Since the pressures in the two systems are the same, i.e.\ $ \Pi
= \rho$, this is equivalent to equating the two insertion
free-energies $F_1$ of Eq.~(\ref{eq13.2}) and Eq.~(\ref{eq13.4}), as
done in references\cite{Meij94,Eise96}.  For $q=R_g/R_c \rightarrow 0$
this expression reduces to $R_{AO}^{eff}/R_g = 2/\sqrt{\pi}$, the
known result for a single wall.  For large $q$ on the other hand hand,
the effective AO radius scales as $R_{AO}^{eff}/R_g \sim q^{-1/3}$.
For fixed $R_g$, the effective radius $R_{AO}^{eff}$ decreases
monotonically with decreasing sphere size $R_c$, as shown in
Fig.~\ref{fig:EJM-Fig2}.

\subsection{Interacting polymers}

\subsubsection{Low density limit for interacting polymers}
 For interacting polymers, the $\rho \rightarrow 0$ limit of the
curvature corrections to the surface tension have been calculated to
first order in an $\epsilon$-expansion by Hanke, Eisenriegler, and
Dietrich\protect\cite{Hank99}.  For large spheres (small $R_g/R_c$),
they find:
\begin{eqnarray}\label{eq13.7}
\lim_{\rho \rightarrow 0} \frac{\gamma_s(\rho)}{\gamma_w(\rho)} &
  \approx & 1 + 0.849 q \nonumber \\ & -& 0.0375 q^2 + {\cal O}(q^3)
\end{eqnarray}
  In this low density and small $q$ regime, the
curvature corrections for interacting polymers are quite similar to those
found for non-interacting polymers. Compare, for example, the first
relative curvature correction coefficient, which is $0.849$ for interacting
polymers, and $0.886$ for ideal polymers.  In the opposite (small
sphere) $q \rightarrow \infty$ limit the differences are more
pronounced: $\gamma_s^{id}/\gamma_w^{id} \sim q$,
$\gamma_s^{AO}/\gamma_w^{AO} \sim (q_{AO})^2$ while for
interacting polymers RG and scaling theory approaches predict that
$\gamma_s/\gamma_w \sim (q)^{1/\nu-1}\approx
q^{0.701}$\cite{deGe79a,Hank99,Eise00}.

\subsubsection{Interacting polymers at finite densities}

\begin{figure}
\begin{center}
\epsfig{figure=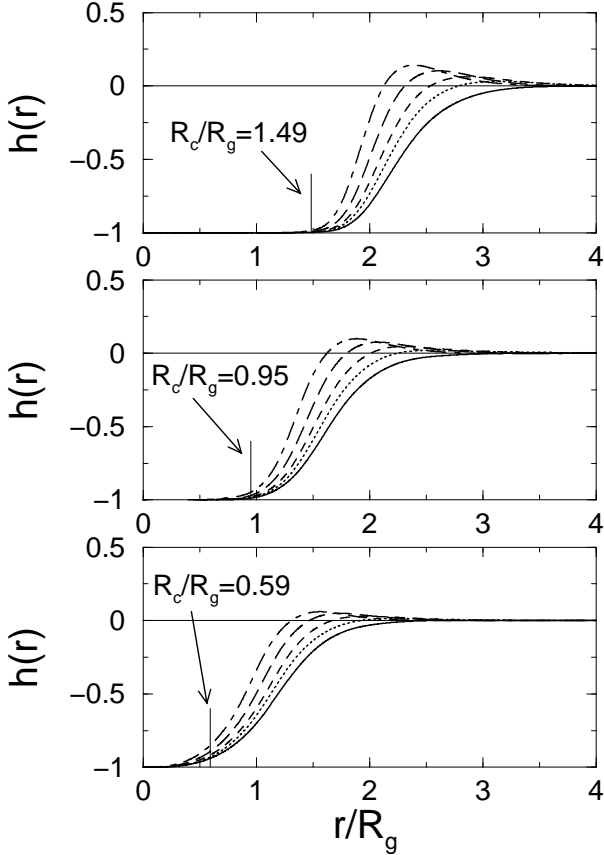,width=8cm}
\caption{\label{fig:hrsphereall} The polymer CM density profile $h(r)$
around a sphere for the ratios $q=R_g/R_c = 0.67, 1.05, 1.68$ (graphs
from top to bottom).  For each sphere size the curves are for
$\rho/\rho* = 0.037,0.30,0.59,1.17,2.33$ (solid, dotted, dashed,
long-dashed, and dot-dashed lines respectively).  The depletion layer
narrows with increasing density, just as was found for as single wall
(compare with Fig.~\protect\ref{fig:hz}).  The small vertical lines
denote the position of the radius of the colloid.  The polymers can wrap
more easily around the smaller colloids, which explains why the CM
profile penetrates further into the colloid for smaller $R_c/R_g$.}
\end{center}
\end{figure}

We have calculated the density profiles $h(r) = \rho(r)/\rho -1$ for
polymers around spheres of radius $R_c = 1.49 R_g$, $R_c = 0.95 R_g$
and $R_c = 0.59 R_c$ from simulations of $L=500$ SAW polymers.  These
are shown in Fig~\ref{fig:hrsphereall} in the CM representation.  Just
as was found for the case of a planar wall, the depletion layers
shrink with increasing bulk density $\rho$.  Because the polymers can
deform around the colloid, the density profiles in the CM
representation can penetrate into the HS region, an effect which
becomes more pronounced for smaller colloids (larger $q$).
(For an interesting proposal that describes the monomer density around 
a spherical particle we refer to reference\cite{Tuin01c}.)

The relative adsorption around a sphere is defined as:
\begin{equation}\label{eq13.21}
4 \pi R_c^2\hat{\Gamma}_s(\rho) = - \frac{4 \pi
 R_c^2}{\rho}\frac{\partial (\Omega^{ex}/A)}{\partial \mu} =
 \int_{0}^\infty 4 \pi r^2 h(r) dr + \frac43 \pi R_c^3.
\end{equation}
Here $h(r)$ is defined from the center of the sphere, and
$\hat{\Gamma}_s(\rho)$ has the dimension of a length.  The volume of a the
single HS was subtracted off so that the adsorption only describes the
effects of the depletion layer around a sphere. For low density the
relative adsorption of a sphere is larger than that of a planar wall
by a curvature correction factor  term similar to those described in 
Eq.~(\ref{eq13.7}).  As the density increases the relative adsorption
decreases and  tends asymptotically to the same value as for a
planar wall. This can be understood from the simple ``blob''
picture\cite{deGe79} in the semi-dilute regime: Since the ratio of the
blob-size to the sphere $\xi(\rho)/R_c$ decreases with increasing
density, the curvature corrections to the relative adsorption are also
expected to become relatively less important with increasing density.

\begin{figure}
\begin{center}
\epsfig{figure=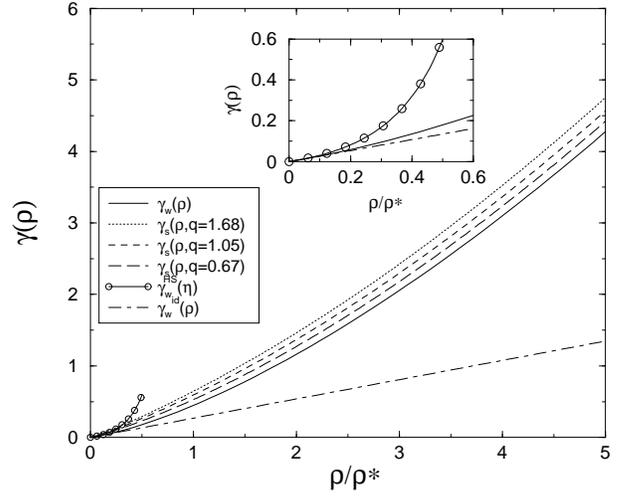,width=8cm}
\caption{\label{fig:gammas} Surface tension for a planar wall, and for
spheres with $q=0.67$, $q = 1.05$ and $q = 1.68$ as a function of
density.  We also include the planar surface tension of a HS fluid,
with $R_p=R_g$ such that $\eta_p=\rho/\rho^*$.  Inset: Blowup of the
graph for low  densities.  The planar wall surface tension for
interacting polymers, ideal polymers, and the HS system are compared.  }
\end{center}
\end{figure}

\begin{figure}
\begin{center}
\epsfig{figure=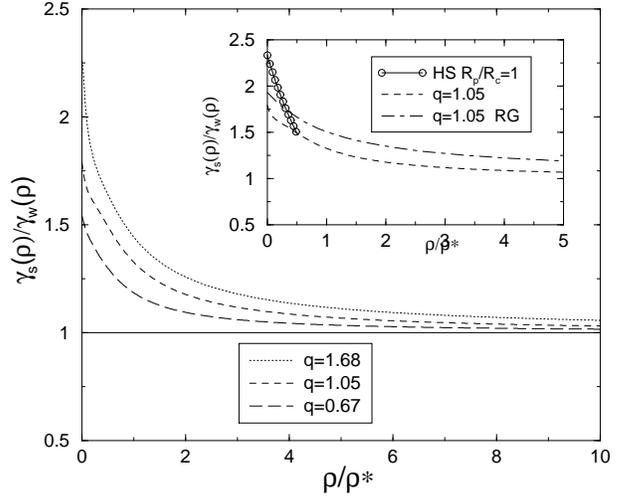,width=8cm}
\caption{\label{fig:ratio-gamma} Ratio of surface tension of a sphere
to the surface tension of a wall for spheres with $q=0.67$, $q = 1.05$
and $q = 1.68$.  Inset: Comparison of an RG
calculation\protect\cite{Maas01} valid to lowest order in $q$, and our
direct calculation for $q=1.05$. We also compare the ratio of the
surface tension of a HS fluid around a single inserted sphere to the
planar HS surface tension.  The size-ratio is $1:1$, and
$\eta_p=\rho/\rho^*$. The value at $\eta_p=0$ is equal to that of the
AO model, given by Eq.~(\protect\ref{eq13.5}). }
\end{center}
\end{figure}

The surface tension $\gamma_s(\rho)$ can now be calculated from
Eq.~(\ref{eq2.2}) using the adsorption defined in Eq.~(\ref{eq13.21}).
In Fig.~\ref{fig:gammas}, we compare the surface tension for three
different sphere sizes to $\gamma_w(\rho)$, the value for a planar
wall.  As expected from the results for low densities, (see e.g.\
Eq.~(\ref{eq13.7})), for a given density $\rho$, the surface tension
increases with decreasing $R_c$.  The ratio
$\gamma_s(\rho)/\gamma_w(\rho)$, shown in Fig.~\ref{fig:ratio-gamma}
decreases with increasing $\rho/\rho^*$.  Again, the rate of change is
largest in the dilute regime; for increasing $\rho/\rho^*$ the two
terms appear to approach each other asymptotically.  Just as we argued
for the adsorptions, the ``blob'' picture in the semi-dilute regime
implies that the curvature corrections should decrease with increasing
density, which is what we observe.  This also implies that
$\gamma_s(\rho) \approx \gamma_w(\rho) \sim \rho^{1.539}$ in the
semi-dilute regime.  Of course the smaller the HS, the higher the
density one needs for the curvature corrections to become
negligible. This picture is confirmed by recent scaling and RG
arguments\cite{Maas01}, which show that the first curvature correction
coefficient $\kappa_1 \sim \rho^{\nu/(3 \nu -1)}\sim \rho^{0.770}$,
implying that with increasing density, the contribution of the
curvature corrections defined in Eq.~(\ref{eq1.2}) becomes relatively
smaller, so that $\gamma_s$ approaches $\gamma_w$.  In the inset of
Fig.~\ref{fig:ratio-gamma} we compare our results to the RG
calculations, valid to lowest order in $q$, i.e. $\gamma_s(\rho) =
\gamma_w(\rho) + \kappa_1(\rho) q$.  Although only the results for the
ratio $q=1.05$ are shown, they are similar to those at the other two
size-ratios, which also show an overestimate by the RG.  The
difference may be due in part to higher order $\kappa_i(\rho)$ terms
which have not yet been calculated by RG.  To confirm this picture
further simulations are needed since: (a) Our simulations of the
adsorption are only for $\rho/\rho^* \leq 2.32$, and we extrapolated
to higher densities using a fit form which scales as
$-\hat{\Gamma}_s(\rho) \sim \xi(\rho) \sim \rho^{-0.770}$ at high
densities. (b) We only examined $3$ different sphere sizes so that it
is difficult to directly extract $\kappa_1(\rho)$, and for that matter
the higher order $\kappa_i(\rho)$.

Finally, we reemphasize how much the density dependence of the surface
tension of the interacting polymers differs from that of ideal
polymers or the related Asakura-Oosawa model, where the ratio of the
wall to the sphere surface tensions is independent of density, and
close to that of interacting polymers in the low-density
limit (Compare Eqs.~(\ref{eq13.3}),~(\ref{eq13.5}), and
(\ref{eq13.7})).

\subsubsection{Comparison with a hard sphere model}

One might inquire what would happen if the polymers were modeled as HS
instead.  By using the very accurate Rosenfeld fundamental measure
density functional\cite{Rose89} technique, an explicit form for the
surface tension of a HS fluid with radius $R_p$ around a single HS
(radius $R_c$) has been calculated\cite{Roth01}
\begin{eqnarray}\label{eq13.7a}
\frac{\gamma_s^{HS}(\eta_p)}{\gamma_w(\eta_p)} & = & 1 + \frac{2
 (1-\eta_p)}{(2 + \eta_p)} (\frac{R_p}{R_c}) \nonumber \\ &- & \frac{2
 (1-\eta_p)^2 \ln(1-\eta_p)}{3 \eta_p (2+\eta_p)} (\frac{R_p}{R_c})^2 ,
\end{eqnarray}
where we have defined the packing fraction $\eta_p = \frac43 \pi
\rho_p R_p^3$, for a number density $\rho_p$.  The planar wall-surface
tension is given by
\begin{equation}\label{eq13.7b}
 \gamma_w(\eta_p) = \frac{3 \eta_p (2 + \eta_p)}{8 \pi R_p^2 (1 - \eta_p)^2}.
\end{equation}
We note that this result for the surface tension of a HS fluid around
a sphere was also derived independently by scaled particle
theory\cite{Hend83}. Eq~(\ref{eq13.7a}) can be generalized to the
non-additive HS model, for which the cross diameter $R_{cp} \neq
\frac12(R_c + R_p)$, so that one can smoothly interpolate between the
fully repulsive HS model and the fully non-additive AO
model\cite{Roth01,nonadd}.

To lowest order in density, the surface tension of a HS system near a
planar wall is $ \gamma_w^{HS} \approx R_p \rho$, i.e. the same as
that of the AO model, which is close to that of interacting polymers
in the same limit where $ \gamma_w \approx 1.074 R_g \rho$. However,
the terms of order $\rho^2$ are already significantly larger in the HS
case. Therefore, as illustrated in the inset of Fig~\ref{fig:gammas},
the HS model gives a large relative overestimate of the surface
tension well before reaching the packing fraction at which the system
freezes. (here we took units where $R_p=R_g$ so that $\eta_p$ is
equivalent to $\rho/\rho^*$).

For a fixed size-ratio $R_p/R_c$ the curvature corrections for a HS
system vary with density  as:
\begin{equation}\label{eq13.23}
\frac{\gamma_s^{HS}(\eta_p)}{\gamma_w^{HS}(\eta_p)} =
\frac{\gamma_s^{AO}}{\gamma_w^{AO}} - \left(\frac{3 R_p}{2 R_c}  
+ \frac{2 R_p^2}{3 R_c^2} \right) \eta_p  + {\cal O}(\eta_p^2)
\end{equation}
where the ratio for the AO model comes from Eq.~(\ref{eq13.5}) with
$R_{AO}=R_p$.  As illustrated for a 1:1 size ratio in the inset of
Fig.~\ref{fig:ratio-gamma}, for small $\eta_p$ this ratio is indeed
almost linear.  The change with density is  more pronounced than that
found for a polymer-colloid system with a similar size-ratio,
suggesting (not surprisingly) that a full HS system is not such a good
model of interacting polymers, even at relatively low densities.
Making the spheres non-additive does not fundamentally alter this
picture -- the behavior of polymers falls into the class of ``mean
field fluids''\cite{Loui00a,Loui01a} i.e.\ they do not behave like hard-core
particles.

\subsubsection{Direct calculation of $F_1$ by the  Widom insertion method}

We also performed direct computer simulations of the free energy
$ F_1$ by measuring the insertion probability of a single sphere
in a bath of polymers at fixed density $\rho/\rho^* = 1.16$.  This is
closely related to the so-called Widom insertion technique to find the
chemical potential\cite{frenkelbook}. Fig.~\ref{fig:chempot} shows
that $ F_1$ grows with increasing sphere size as expected.  The
same is true for the contribution due to the depletion layer, i.e.\
the contribution proportional to $4 \pi \gamma_s R_c^2$ in
Eq.~(\ref{eq1.1}).  However, the relative importance of this surface
tension term increases with {\em decreasing} sphere size, and becomes
the dominant contribution as $R_c/R_g \rightarrow 0$.  The values up
to $R_c/R_g =0.59$ were calculated by the insertion probability
method, while those with larger $R_c/R_g$ were taken from the
adsorption method, i.e. from the density profiles, as was done for
example in Fig.~\ref{fig:gammas}.  For $R_c/R_g = 0.59$ we used both
methods and find very similar results, suggesting that the two
approaches are mutually consistent.  We also compare to results
for ideal polymers (Eq.~\ref{eq13.2}) and for PRISM\cite{Fuch01}.

\subsubsection{Limit of small colloids}

\begin{figure}
\begin{center}
\epsfig{figure=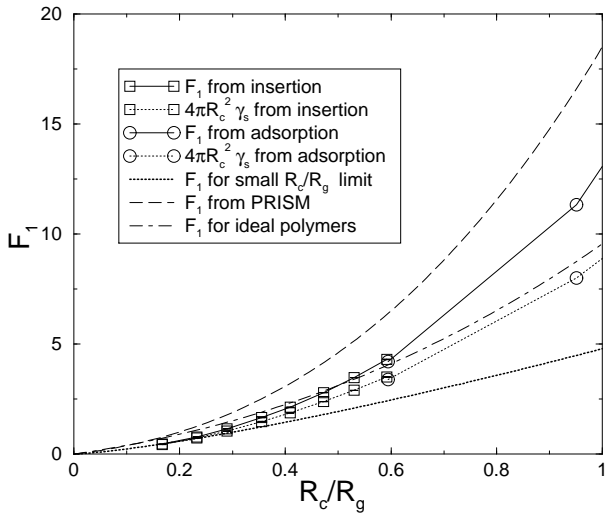,width=8cm}
\caption{\label{fig:chempot} Insertion free energy $F_1$ for spheres
of various radii $R_c$, in a polymer bath at $\rho/\rho^* = 1.16$.
For smaller $R_c/R_g$ a direct insertion method was used, while for
larger $R_c/R_g$ the adsorption method was used. We also compare $4
\pi R_c^2 \gamma_s(\rho)$, the contribution to $ F_1$ due to the
creation of a depletion layer.  For small $R_c/R_g$ this term is the
dominant contribution to the insertion free energy $F_1$.  Comparison
is also made to an expression from RG theory, Eq.~(\protect\ref{eq13.8}),
valid in the small $R_c/R_g$ limit\protect\cite{Eise00}, with  results
from the PRISM approach\protect\cite{Fuch01}  and with $F_1$ for ideal
polymers, taken from Eq.~(\protect\ref{eq13.2}).  Note that for this
density, the ideal and interacting results for $ F_1$ cross each other
at $R_c/R_g \approx 0.5$, below which it is easier to insert a
spherical colloid into an interacting polymer solution than into a
non-interacting one.}
\end{center}
\end{figure}

\begin{figure}
\begin{center}
\epsfig{figure=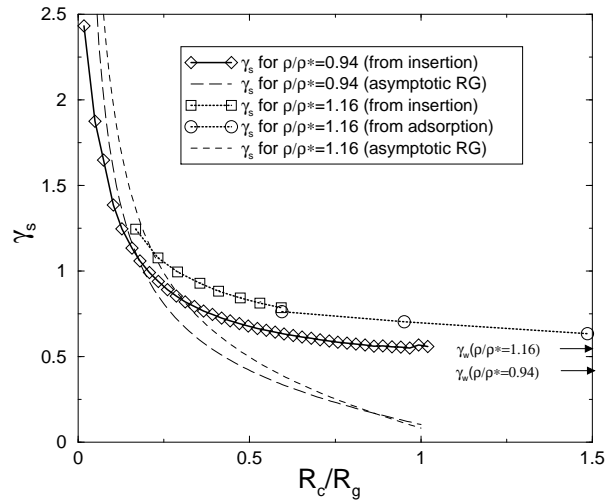,width=8cm}
\vspace*{1cm}
\caption{\label{fig:L2000gamma} Surface tension $\gamma_s$ for spheres
of different radius $R_c$, in a polymer bath at $\rho/\rho^* = 1.16$,
($L=500$ SAW simulations) and for $\rho/\rho* =0.94$ ($L=2000$ SAW
simulations). The insertion and the adsorption methods agree to within
the expected statistical errors of our approach for $R_c = 0.59 R_g$.
We also compare to an expression from RG theory,
Eq.~(\protect\ref{eq13.8}), valid in the small $R_c/R_g$
limit\protect\cite{Eise00}.  The arrows on the right depict the values
of the planar surface tensions, valid as $R_c/R_g \rightarrow
\infty$. }
\end{center}
\end{figure}

In the limit of small $R_c/R_g$, scaling arguments and RG theories
 predict that the free energy to insert a single particle in a bulk
 polymer solution takes the form\cite{Eise00,deGe79a}
\begin{equation}\label{eq13.8} 
 F_1 = A_g R_c^{d-1/\nu} \rho R_g^{1/\nu}
\end{equation}
Where $A_g$ is a universal numerical pre-factor that can be calculated
from an RG technique\cite{Hank99,Eise00}.  For $d=3$
Eq.~(\ref{eq13.8}) reduces to $ F_1 \approx 18.4 \rho R_c^{1.30}
R_g^{1.70}$. This expression is directly compared to our simulations
in Fig.~\ref{fig:chempot}.  By comparing to Eq.~\ref{eq1.1} we can
extract the surface tension from the insertion free energies.  This
was done for the $ F$ at $\rho/\rho*=1.16$ shown
Fig.~\ref{fig:chempot}, and also for $L=2000$ polymers at
$\rho/\rho*=0.94$.  Using the longer polymers allows effectively
smaller colloidal $R_c$'s to be used in our lattice simulations.  The
surface tensions are depicted in Fig.~\ref{fig:L2000gamma}.  At small
$q$ our computer simulation results correspond reasonably well with
the asymptotic RG results.  We expect there to be small errors due to
the discreteness of out lattice simulations, similar to those depicted
for ideal polymers in Fig.~\ref{fig:EJM-Fig2}.  These discretization
errors become more important as the spheres become relatively
smaller. We have made some small corrections\cite{corrections} to take
this into account, but a more systematic study, possibly with longer
polymers, would be necessary to completely test the RG results.

When the colloids are much smaller than the polymers, one expects that
they only probe the local monomer density, and not the overall number
density of polymer coils. In fact, Eq.~(\ref{eq13.8}) implies just that
since $F_1 \propto \rho R_g^{1/\nu} R_c^{1.30} \sim \rho L R_c^{1.30} =
c R_c^{1.30}$.  The reason $F_1$ scales linearly with the monomer
density $c$ is that by definition this is very small ($c \ll 1$) in
the dilute and semi-dilute regime. The small colloidal particles
probe what is effectively an ideal gas of monomers.

For ideal polymers $F_1 \propto c R_c$ in the limit of small $R_c$,
which implies that for a given $R_g$ and $\rho$, and for a small
enough $R_c$ it is easier to insert a hard-sphere into a bath of
interacting polymers than it is to insert it into a bath of
non-interacting polymers. At first sight this may seem surprising, but
the reason is as follows: Inside an interacting polymer, the monomer
concentration scales as $c \sim R_g^{-1.30}$ while for ideal polymers
it scales as $c \sim R_g^{-1}$.  In other words, the interactions
swell a polymer and make it less dense; for a given $R_g$, the monomer
density $c$ is larger for ideal polymers than for interaction
polymers, and since the small colloids only probe the local monomer
density it is easier to insert the sphere into an interacting system
than into a non-interacting system at the same $R_g$.  This effect is
illustrated in Fig.~\ref{fig:chempot} for $\rho/\rho* = 1.16$, where
the crossover is at about $R_c \approx 0.5 R_g$. (This limit should
not be confused with a comparison at fixed $L$).  Note that the PRISM
results also overestimate $ F_1$ at small $R_c$.  This is in part
because the simplified PRISM model we compare to also includes ideal
polymer statistics, resulting in an overestimate of the monomer
density compared to a true interacting system, an effect already
pointed out in ref.\cite{Fuch01}.

For large spheres, on the other hand, where $F_1 \approx \frac{4}{3}
\pi \Pi(\rho) R_c^3$ which scales as $F_1 \sim \rho^{2.30} R_c^3$ in
the semi-dilute regime, the spheres do directly probe the number
density of polymer coils, and the insertion free energy for
interacting polymers is always higher than that of ideal polymers at
the same $R_g$ and $\rho$.  Note how differently the large and small
$R_c/R_g$ limits of $F_1$ scale both with $\rho$ and with $R_c$.
Significant differences can be also found for the scaling of the
surface tensions since for large $R_c/R_g$, $\gamma_s \approx
\gamma_w(\rho)(1 + {\cal O}(R_g/R_c)) \sim \rho^{1.539}(1 + {\cal
O}(\rho^{-0.770}R_g/R_c)$, while for small $R_c/R_g$, the RG
expressions imply that $\gamma_s \sim c R_c^{-0.7}$.

\subsection{Surface tension for polymers as soft colloids}

We have recently modeled polymers as single ``soft colloids''
interacting with a pair potential between their
CM\cite{Loui00,Bolh01,Bolh01a,Bolh01b}. These pair potentials were
derived by a liquid state theory based inversion procedure such that the
soft colloids have exactly the same radial distribution function
$g(r)$ as those generated by a fully interacting polymer simulation.  A
similar inversion procedure was used to derive the potential between the
soft-colloids and a planar wall or a HS.  These wall-polymer or
sphere-polymer potentials are such that they exactly reproduce the
one-body density profiles $\rho(r)$.

Since our effective polymer-polymer potentials provide a very accurate
representation of the pressure $\Pi(\rho)$\cite{Bolh01,Bolh01b}, while
the polymer-wall or polymer-sphere interactions are constrained to
reproduce the correct density profiles, and therefore the correct
adsorption $\hat{\Gamma}(\rho)$, Eq.~(\ref{eq2.2}) implies that our
soft-colloid approach has the correct surface tensions automatically
built in.  Similarly Eq.~(\ref{eq1.1}) implies that this approach
correctly reproduces $ F_1$ for a sphere immersed in a polymer
solution.

\section{Conclusions}
In summary then, we have used computer simulations of SAW polymers
on a cubic lattice to calculate the density profiles for non-adsorbing
polymers near a planar wall, and near HS's.  From this we were able to
calculate and fit the relative adsorption $\hat{\Gamma}(\rho)$.
Together with the equation of state, which is well understood for
polymer solutions, this provides the needed ingredients to calculate
the surface tensions through Eq.~(\ref{eq2.2}).

The surface tension of interacting polymers near a planar wall was
shown to differ significantly from that of ideal polymers, or other
simple models such as the Asakura Oosawa penetrable-sphere model, or a
pure HS fluid.  Similarly, a recent PRISM calculation\cite{Fuch01}
also shows large qualitative differences with our results, which could
have been anticipated in view of its use of simplified ideal polymer
statistics.  On the other hand, some recent RG results\cite{Maas01}
compare very well to our calculations.  In the semi-dilute regime, the
surface tension simplifies to the form given in Eq.~(\ref{eq2.7}),
which implies that $\gamma_w(\rho) \propto \xi^{-2} \sim
\rho^{1.539}$.

 Near a sphere with a radius of the same order or larger than $R_g$,
the surface tension $\gamma_s(\rho)$ of the polymer solution can be
written in an expansion in the size-ratio $q$.  For decreasing sphere
size (increasing $q$), the ratio $\gamma_s(\rho)/\gamma_w(\rho)$
increases.  For a given $q$, however, $\gamma_s(\rho)$ approaches
$\gamma_w(\rho)$ as the density increases.  We attribute this to the
decrease of the effective curvature corrections with increasing
density since the blob size scales as $\xi(\rho)\sim \rho^{-0.770}$ in
the semi-dilute regime. This is again consistent with some recent RG
calculations of the correction coefficient
$\kappa_1(\rho)$\cite{Maas01}, although further simulations are needed
to confirm the scaling and form of the coefficients $\kappa_i(\rho)$.

For smaller colloids, it is advantageous to use a direct Widom
insertion technique to calculate the free-energy $ F_1$. For very
small colloids (large $q$), our simulations were consistent with
asymptotic RG results which suggest that $ F_1 \propto c
R^{1.30}$.  This insertion free-energy is dominated by the
contribution of the depletion layer; its behavior is qualitatively
different from the behavior found at smaller $q$, and suggests that
the expansion of Eq.~(\ref{eq2.2}) breaks down for large $q$.

Because our soft-colloid approach was derived to reproduce the correct
one-particle density profiles near hard walls or hard-spheres, it will
automatically reproduce the correct adsorptions, and therefore also
the correct surface tensions and insertion free-energies $ F_1$.

The walls and spheres in this study are purely repulsive.  Adding a
wall-polymer attraction should decrease the amount of depletion, and
therefore also lower the surface tensions.  More subtle effects could
be expected if in addition the solvent quality is decreased.  New
effects are also expected for binary mixtures of polymers with
selective adsorption of one of the species.  These systems will be the
subject of future investigations.

The next step is to move from the one-sphere or one-wall problem to
the case of a two-sphere or a two-wall system, and calculate the
effective interactions between the two particles.  This is the subject
of a forthcoming paper\cite{Loui01d}.

\section*{Acknowledgements}
 AAL acknowledges support from the Isaac Newton Trust, Cambridge, and
the hospitality of Lyd\'{e}ric Bocquet at the Ecole Normale Superieure
in Lyon, where much of this work was carried out.  PB acknowledges
support from the EPSRC under grant number GR$/$M88839, EJM
acknowledges support from the Royal Netherlands Academy of Arts and
Sciences.  We thank L. Bocquet, R. Evans, L. Harnau, and H. L\"{o}wen
for helpful discussions.  We thank R. Tuinier and H.N.W. Lekkerekerker
for sending us their preprints\cite{Tuin01,Tuin01c} prior to
publication, and E. Eisenriegler for sending us numerical results
from\cite{Maas01}.  V. Krakoviack is thanked for pointing out the 
small error in $R_g$\cite{Rg}.


\newpage

\section*{List of figures}	

\end{multicols}


\begin{thebibliography}{99}
\bibitem{gamma0} To compare with \protect\cite{Hank99}, who expressed
their results in terms of $R_e$, the end-to-end radius, we have used
$R_g \approx 0.406 R_e$, the value expected for interacting
polymers\protect\cite{Doi86}.  This is close to the ideal-polymer
result $R_g = 1/\sqrt{6} R_e \approx 0.408 R_e$.
\bibitem{nonadd} To correctly obtain the fully non-additive AO limit,
the explicit $R_p$ in Eq.~(\protect\ref{eq13.7a}) should be equated
with $R_{AO}$, while the packing fractions $\eta_p \rightarrow 0$.
\bibitem{corrections}In the simulation we insert the colloid with its
center a lattice site.  We then check all the sites that are within
the radius for overlap with one of the polymers. The number of these
lattice sites, i.e. the volume occupied by the sphere can be
transformed into an effective radius. This radius takes the
discreteness effects partially into account, and is used in the
analysis.
\end{thebibliography}
\end{document}